\documentclass[manuscript,noblind]{geophysics}
\usepackage[T1]{fontenc}
\usepackage{float}
\usepackage[hidelinks]{hyperref}
\renewcommand{\figdir}{figures}
\begin{document}
\title{FA-Agent: An extensible multimodal agent framework for first-arrival picking in mountainous seismic data}
\renewcommand{\thefootnote}{\fnsymbol{footnote}}
\address{
\footnotemark[1]Earth Observatory of Singapore, \\
Nanyang Technological University, Singapore \\
\footnotemark[2]School of Civil and Environmental Engineering, \\
Nanyang Technological University, Singapore \\
\footnotemark[3]Corresponding author. E-mail: congcong.yuan@ntu.edu.sg (C. Yuan) \\[6pt]
Preprint of a manuscript submitted to GEOPHYSICS (under review)
}
\author{Wenda Li\footnotemark[1] and Congcong Yuan\footnotemark[2]\footnotemark[3]}
\lefthead{Li \& Yuan}
\footer{Agent framework for first-arrival picking}
\righthead{Agent framework for first-arrival picking}
\maketitle

\begin{abstract}
First-arrival traveltimes underpin near-surface velocity modeling and static corrections, but in mountainous land data with a low signal-to-noise ratio the first arrival is often weak, intermittent, and masked by strong direct waves and noise. Energy-based pickers then pick later arrivals or noise, and deep-learning pickers usually require retraining for a new survey. Because erroneous picks bias first-arrival tomography, many difficult zones are still inspected and corrected manually. We present FA-Agent, an agent framework for first-arrival picking built on a multimodal large language model. Like a software package, it combines a model that performs visual recognition and makes decisions, tools that process the data, and skills that carry the experience of processing mountainous data. The framework is open and extensible: new pickers, denoising methods, and quality-control tools can be added, and new experience can be written into the skills. It is restricted neither to mountainous data nor to a particular model. Emulating a seismic data processor, the agent first surveys the whole gather to find zones that need attention, then examines them in enlarged views, and finally keeps, corrects, or removes the initial picks. On semi-synthetic gathers with known first arrivals and real mountainous noise, the agent picked the first arrival with high accuracy where it was clearly visible and removed the picks where the first arrival was invisible and could not be judged. Applied to randomly sampled gathers from a mountainous survey, it corrected picks onto weak first-arrival branches and removed picks where no first arrival could be confirmed, and it performed well on the most difficult data. Cases reviewed by experts are added to the skills automatically, so the agent improves with use, and a locally deployed model keeps confidential data on site. FA-Agent can thus become an essential component of future automated seismic data processing workflows.

\end{abstract}

\section{Introduction}
\label{sec:introduction}

First-arrival traveltimes constrain near-surface velocity models and static corrections in land seismic processing \citep{Zhu1992,Yilmaz2001}. The accuracy with which a first arrival can be timed depends on its signal-to-noise ratio, and the visible onset depends on the noise and the display gain \citep{Hagedoorn1964,Hatherly1982}. Automatic pickers exploit short-term to long-term average ratios \citep{Allen1978}, onset estimators based on the Akaike information criterion (AIC) \citep{Maeda1985,Sleeman1999}, correlation and statistical criteria \citep{Gelchinsky1983}, and waveform correlation \citep{Molyneux1999}. Other pickers use energy ratios \citep{Coppens1985}, fractal dimension and entropy \citep{Boschetti1996,Sabbione2010}, image segmentation of energy-ratio records \citep{Mousa2011}, and instantaneous traveltime \citep{Saragiotis2013}. Weak first arrivals can be enhanced before picking, for example by the $\tau$-$p$ transform of energy-ratio records \citep{Mousa2012} or by interferometry \citep{Place2019}. Gather-level consistency constraints further improve these picks and reject erroneous arrivals \citep{Sabbione2010}, and global path tracing enforces the stability and smoothness of the picked curve \citep{ZhangD2022}. Interactive and assisted schemes reduce, but do not remove, the manual effort \citep{Giroux2009,AlTheyab2016}. Nevertheless, picks in zones with weak onsets or interfering noise still require visual inspection and manual editing. In large three-dimensional surveys, even a small fraction of mispicked traces can consume substantial operator time.

Neural networks were applied to first-break picking and trace editing early on \citep{McCormack1993}, and deep learning now addresses part of this problem by learning the spatial and temporal patterns of arrivals from labeled records. Applications have progressed from convolutional waveform classification \citep{Yuan2018} and combined convolutional and recurrent networks \citep{YuanP2020} to U-net and pixel-wise segmentation \citep{Hu2019,Ma2020}, multistage segmentation \citep{Wang2024}, and three-dimensional U-net picking with waveform and energy-semblance inputs \citep{Han2022}. Self-training from single to multiple data domains \citep{Ozawa2024} and refinement of the predicted arrival times \citep{Pu2024} further improve the picks. Similar methods are used to pick arrival times of microseismic events \citep{Song2010,Tan2016,Wu2019,ZhangG2020,Zheng2020,Cano2021,ZhangX2021,Sheng2022}, for which \citet{Akram2016} reviewed the picking algorithms. For mountainous data, \citet{LiW2024} incorporated attention gates and energy-semblance information to improve picking under strong noise and missing or bad traces. Performance on a new survey can nevertheless depend on differences in acquisition conditions, waveforms, and noise \citep{StCharles2024,Wang2024}; fine-tuning and transfer learning reduce this mismatch \citep{Mardan2024,LiH2024}, but require adaptation data. Hybrid methods can also detect and correct mispicks with a neural network \citep{Duan2020}, and a U-Net can correct first-arrival traveltimes obtained by refraction interferometry \citep{Lee2024}. Even so, these methods are designed mainly to predict arrival times. Whether a pick in a difficult zone is supported well enough to be used is usually decided later, in quality control and manual review.

What remains difficult to automate is therefore a decision rather than a prediction: whether the data support a given pick. This decision matters because of how first arrivals are used. Traveltime tomography combines many overlapping raypaths, and least-squares inversion is sensitive to erratic data \citep{ClaerboutMuir1973}; arrival-time uncertainties are used to weight the data in inversion \citep{Abakumov2020}. An erroneous pick can bias the velocity model, whereas a missing pick removes only one of many overlapping constraints. For zones where the arrival is difficult to judge, we follow a conservative principle and remove the pick rather than keep a doubtful one. In the mountainous gathers considered here, such decisions are frequent. First arrivals may be weak or masked by noise, and along far-offset branches visible and obscured segments often alternate, forming a pattern we refer to as \emph{Christmas-tree-like} (Figure~\ref{fig:fig1_0813_challenges}a).
A processing geophysicist surveys the whole gather, enlarges questionable zones, and assesses each onset from waveform changes and neighboring arrivals. Local slope can guide the search for a faint arrival but cannot establish an arrival where no signal is visible. Each pick is then kept, adjusted, or removed.

With multimodal large language models (LLMs) and LLM agents, such visual patterns can be recognized directly from gather displays and linked to processing actions. Visual instruction tuning enables a model to interpret images under written guidance \citep{Liu2023}, and agent frameworks let a model select tools and use their outputs in later decisions \citep{Yao2023}; such multistep tool use has been demonstrated in chemistry \citep{Bran2024,Boiko2023}. In geophysics, LLMs and large vision and multimodal models have been reviewed as emerging tools for exploration tasks \citep{LiuQ2025}. LLM agents have been used to select and execute seismic processing tools from natural-language requests \citep{Kanfar2025}, to automate seismic wave simulation \citep{Ren2025}, and to conduct multistep seismological analyses with auditable evidence chains \citep{LiuF2026}. \citet{Hennenfent2026} further embedded expert judgment in agents that diagnose and attenuate seismic noise. These agents mainly orchestrate processing, modeling, or analysis workflows. To our knowledge, using a multimodal model's visual judgment to decide, window by window, whether first-arrival picks should be kept, corrected, or removed has not been examined. Such use also requires explicit guidance on how a processing geophysicist evaluates ambiguous arrivals. In current agent frameworks, this guidance can be packaged as \textit{skills}: reusable instructions and reference files that an agent loads when a task requires them. Expert review criteria and reviewed examples can thus be supplied to the agent and revised without changing model weights.

We propose \textit{FA-Agent}, an agent framework for first-arrival picking on difficult data that combines conventional processing, visual recognition, and expert experience. Conventional picking provides candidate arrivals across the gather. The core of the agent is visual recognition: it reads the gather display as a processing geophysicist does, identifies zones where the candidates depart from the visible onset, and examines enlarged views to keep, correct, or remove the picks. We organized our experience in correcting picks for weak first arrivals, noise-masked onsets, missing traces, and intermittent far-offset branches, together with expert experience, into skills that guide these visual decisions. Each decision is recorded with its visual evidence for review. We first test FA-Agent on semi-synthetic gathers whose first arrivals are known, and then apply it to mountainous seismic data with a low signal-to-noise ratio, where it obtains good results. The agent processed 1000 randomly sampled gathers; three of them, all outside its case library and two from the most difficult class, are presented here. The following sections describe the agent design, present the semi-synthetic tests and examples from mountainous field data, and discuss the framework and future work. FA-Agent is expected to substantially reduce the manual work of first-arrival picking.

\section{Method}
\label{sec:method}

\subsection{Challenges of mountainous data and overall workflow}

First arrivals in mountainous land data are difficult to pick for reasons that differ along a gather, as illustrated by gather 0813 (Figure~\ref{fig:fig1_0813_challenges}a). Rugged topography and a heterogeneous near surface make the first-arrival curve irregular. On the left of the gather, a strong linear event, the direct wave, carries more energy than the first arrival. At these offsets the direct wave travels more slowly than the refracted wave, so the first arrival is the weaker branch that arrives earlier, above it. Bands of vertical noise interrupt the gather, and at far offsets on the right the first arrival is obscured by low-frequency interference. Elsewhere, for example around the apex, the first arrival is clear. The difficulties are concentrated in particular zones, and these are also where a conventional pick fails (Figure~\ref{fig:fig1_0813_challenges}b): it follows the strong event instead of the weak first-arrival branch, crosses a noise band, and continues through the interference where no first arrival is visible.

An experienced processing geophysicist deals with such gathers by first looking at the whole gather to understand the first-arrival pattern and to locate the problem zones, then examining those zones closely, and finally keeping, correcting, or removing the picks there. FA-Agent follows the same sequence (Figure~\ref{fig:fig2_workflow}):
a conventional picker provides an initial pick for every trace, the agent identifies from images of the whole gather the zones that need attention, and it then inspects these zones at higher resolution to correct or remove the picks.

\subsection{FA-Agent components}

FA-Agent consists of a multimodal model, a set of tools, and a set of skills. The model is the decision maker: it reads the task instructions, looks at seismic images, decides what to examine next, and calls tools to carry out its decisions. Tools are local programs that perform deterministic operations. They read SEG-Y data, attenuate anomalous energy with a denoising module, compute the conventional picks, render whole-gather and enlarged views with or without picks, convert visual anchors into traveltimes, and export and check the results. All views are displayed with automatic gain control (400~ms window, maximum gain 30). Skills are written instructions with reference cases that describe how the task is done: the order of the workflow, what to look for in the whole gather, how to examine a questionable zone, and when a pick should be kept, corrected, or removed. The model consults the skills at each stage, renders the view it needs, interprets it, and records its decision before moving on. Because the visual judgment is made by the model and the processing by the tools, each decision can be traced to the image on which it was based. Figure~\ref{fig:fig_agent_decision} illustrates this division of labor for window V1 of gather 0813. It shows the image that the agent examines, the decision that it records, and the picks that the tools derive from this decision. In the recorded decision, each trace segment receives an action, a reason, and anchors. The result for this window is discussed further in the field examples.

In this study the visual review was carried out interactively with the multimodal large language model GPT-6 Astra \citep{OpenAIAstra2026}, which inspected the rendered images and called the local tools; the skills followed the format used in that environment \citep{OpenAISkills2026}. No separate application programming interface (API) requests were issued, so per-call model metadata are not available. FA-Agent is not tied to this model. Other current multimodal LLMs also accept images as input, and their visual capability continues to improve. The workflow, tools, and skills can be used with any model that can view images and call tools.

\subsection{Agent workflow}

\paragraph{Conventional initial picking.}
The initial pick combines an envelope energy ratio \citep{Allen1978} and an AIC onset score \citep{Maeda1985} with a lateral continuity constraint that links the picks across the gather. FA-Agent currently integrates this single conventional picker, which performs well where the first arrival is clear and the noise level is moderate. In mountainous data, however, the continuity constraint can lead the curve onto a strong later event, for example the more energetic direct wave instead of the first arrival, or carry it through zones where no first arrival is visible. The initial pick is not accepted as it is; it is the reference that the agent examines. The agent is not limited to this algorithm: better conventional pickers can be added as tools, so that FA-Agent remains an open package that can be extended and updated over time.

\paragraph{Visual recognition of attention windows.}
The agent first views the whole gather without picks, so that its reading of the first-arrival pattern is not biased by the drawn curve, and identifies the earliest coherent wavefront, its branches, and the noisy or missing traces. It then overlays the initial pick and marks a few attention windows where the pick and the visible wavefront disagree or where the wavefront is uncertain, such as a pick lying on a later event or a far-offset zone obscured by noise. This step relies on the whole-gather context. Whether an event is the first arrival or a later one, and whether a weak segment continues an established branch, can be judged only in relation to the rest of the gather, not from individual traces. The windows also focus the subsequent work on the small part of the gather where the initial pick is doubtful.

\paragraph{Visual correction and removal.}
Each window is examined in enlarged views, first without and then with the initial pick, and with enough surrounding traces to show the local trend. At this scale a weak onset that is barely visible in the whole-gather view can often be followed across neighboring traces, and the agent compares the original data with a denoised version when one is available. The agent then divides the window into segments and makes one decision for each segment. The initial pick is kept if it lies on the visible wavefront, corrected by placing anchors on the visible wavefront if it departs from it, or removed if the first arrival cannot be established. Short interruptions of a wavefront visible on both sides may be bridged, but picks are not extended into zones without visible signal. Removal is preferred to a doubtful pick, for the reasons given in the Introduction; it leaves the final pick blank, while the trace and its initial pick are retained. Finally, the agent reviews the assembled picks on the whole gather and on each window, and resolves jumps at segment joints or picks that have moved onto a later branch.

\subsection{Development of the agent}

The skills were developed on mountainous field data rather than written in advance. We randomly sampled 1000 receiver-line gathers from the survey and selected part of them as development cases. On these cases, the agent carried out the correction in consultation with a geophysicist with many years of experience in processing mountainous seismic data. The expert marked problem zones and judged ambiguous events, and the agent learned from these cases: the corrections and the expert's judgments were summarized as experience with mountainous data and written into the skills, without changes to the model weights. The resulting skills are organized into modules for the overall workflow, the whole-gather recognition of attention windows, the local inspection and correction of a window, and the decision standard for keeping, correcting, or removing picks.

The same mechanism continues after development. During large-scale processing, when an expert reviews a gather and contributes experience, the agent automatically writes the reviewed gather into the case library of the skills as a new case. The agent therefore becomes more experienced, and in this sense smarter, the more it is used.

\plot{fig1_0813_challenges}{width=\textwidth,height=0.68\textheight,keepaspectratio}
{Mountainous gather 0813 (720 traces) displayed with automatic gain control (400~ms window, maximum gain 30). (a) Data without picks, with the features that complicate first-arrival picking. (b) The same gather with the conventional initial pick (green), which follows the strong linear event on the left rather than the weaker, earlier branch above it, and continues through the interference at far offsets on the right where no first arrival is visible.}

\plot{fig2_workflow}{width=\textwidth,height=0.68\textheight,keepaspectratio}
{Workflow of FA-Agent. Gray boxes are tool operations; blue boxes inside the dashed outline are the visual-review stages carried out by the agent; the green box is the output. The components on the right support these stages, and expert review of development cases updates the skills.}

\plot{fig_agent_decision}{width=\textwidth,height=0.64\textheight,keepaspectratio}
{Example of how FA-Agent works, for window V1 of gather 0813. (a) The display examined by the agent, without picks. (b) Excerpt of the recorded decision, giving the action, the reason, and the anchors for each trace segment. (c) Picks produced by the tools from this decision: anchors (circles), the corrected pick interpolated between them (orange), the initial pick (dashed), and removed traces (gray bar).}
\section{Semi-synthetic tests}
\label{sec:semisyn}

\subsection{Test design}

In mountainous field data the true first arrival is unknown, and expert picks of weak or obscured arrivals are themselves subjective, so they cannot serve as an objective reference. We therefore first evaluated FA-Agent on semi-synthetic gathers, in which the first arrival is known exactly but the noise comes from mountainous field data.

The signal is a finite-difference synthetic shot gather with 400 receivers at 10~m spacing and a source near trace 211, resampled to the 4~ms interval and 751 samples of the field data. Its amplitude decreases exponentially with offset, so that far-offset arrivals are weaker than near-offset ones. The true first arrival of each trace is defined on the noise-free gather as the first sample whose absolute amplitude exceeds 1\% of the trace maximum, that is, the earliest onset of the first-arrival wavelet.

The noise is real field noise. From each trace of a field gather, we took only the part recorded well above the first arrival, which contains noise but no first-arrival energy. This part was extended to the full record length by mirroring it in time, and the field traces were assigned to the 400 synthetic traces in their original order. The vertical noise bands, the low-frequency interference, and the trace-to-trace variation of the noise level of the field gather are preserved. A single scale factor sets the noise level for the whole gather. We measure it by the signal-to-noise ratio (S/N): the ratio of the root-mean-square amplitude of the signal just after the true first arrival to that of the noise on the same trace.

Two tests were built in this way. In the first test, the noise is taken from gather 0813 at two levels (Figure~\ref{fig:fig_semisyn_noise}). At the moderate level, the median S/N is 60 near the source but about 2 at offsets beyond 150 traces; at the strong level, the noise is doubled, so that the median far-offset S/N falls to about 1. This test examines picking accuracy under strong noise when the first arrival is still present on every trace.

The second test reproduces the situation that motivates the conservative principle: a first arrival that is weak or cannot be seen at all, with stronger later events below it (Figure~\ref{fig:fig_semisyn_weak}a). Using the same synthetic gather, we weakened only the first-arrival wavelet, the 90~ms following the true onset, and left the later events unchanged. Beyond 80 traces from the source, the first-arrival amplitude decreases gradually to half of its original value at the ends of the gather. In addition, the first arrival was hidden in four blocks of traces by reducing it to 4\% of its original amplitude, which makes it practically invisible under the noise. Two blocks are wide (traces 60--92 and 335--370), and two are short (traces 122--134 and 300--311), both no longer than 20 traces; the purple bars in Figure~\ref{fig:fig_semisyn_weak}a mark them. Following the rules described in the Method section, the first arrival should be removed in the wide blocks, whereas the short blocks, with a visible wavefront on both sides, may be bridged. The noise for this test comes from a different class A field gather (0901), which is not used elsewhere in this paper, at the moderate level.

The agent processed the test gathers with the same workflow, display, and skills as the field data, without knowing the truth or how the gathers were constructed, and the skills were not modified for the tests. Accuracy is expressed as the percentage of picks within $\pm4$, $\pm8$, and $\pm20$~ms (one, two, and five samples) of the true first arrival.

\subsection{Picking accuracy under strong noise}

In the first test (Figure~\ref{fig:fig_semisyn_noise}), the conventional pick follows the first-arrival wavelet near its energy maximum, about 40--50~ms after the true onset, and its continuity constraint flattens the apex into a plateau. As a result, only 2\% of its picks lie within $\pm20$~ms of the truth at either noise level. Because the first arrival remains visible on every trace, the agent kept a pick on every trace but corrected all of them onto the earliest onset of the wavelet. All of its picks at the moderate level and all but one at the strong level lie within $\pm20$~ms of the truth, and about 60\% lie within $\pm8$~ms (Figure~\ref{fig:fig_semisyn_noise}c and \ref{fig:fig_semisyn_noise}d). Doubling the noise changed the accuracy of the agent by only a few percentage points, which is consistent with the agent following the wavefront across noisy traces instead of relying on the amplitude of individual traces.

\subsection{Weak and hidden first arrivals}

In the second test, the weakened first arrival on the flanks no longer dominates the gather, and the continuity constraint carries the conventional pick onto the stronger later events over the whole gather; none of its picks lies within $\pm20$~ms of the truth. The agent recovered the visible first arrival on both flanks and across the apex, removed the picks in the two wide hidden blocks (traces 59--92 and 330--371), and bridged the two short ones (Figure~\ref{fig:fig_semisyn_weak}a). The removed traces coincide with the hidden blocks and extend a few traces into their tapered edges, where the first arrival is also weak. All 324 picks that the agent kept lie within $\pm20$~ms of the truth, and 56\% lie within $\pm8$~ms (Figure~\ref{fig:fig_semisyn_weak}b); none was placed on a later event or in noise.

Figure~\ref{fig:fig_semisyn_weak}c relates these decisions to how visible the first arrival actually is. For this evaluation only, each trace was classified from the known signal and noise by the S/N of its first arrival as visible (S/N $\geq 1.5$; 271 traces), marginal (0.5--1.5; 39 traces), or not visible (S/N $< 0.5$; 90 traces). For each class, the bars show the percentage of traces that the agent picked or removed. The agent picked all of the visible traces and 92\% of the marginal ones. Of the traces whose first arrival is not visible, it removed 81\%. The remaining 19\% are the traces inside the two short hidden blocks, which the agent bridged between the visible wavefronts on either side, as the rules allow; these bridged picks are also within $\pm20$~ms of the truth. None of the 76 removed traces had a visible first arrival. The agent therefore picked where the first arrival could be seen, removed it where it could not, and did not replace a hidden first arrival by a later event, which is the behavior required by the conservative principle.

\plot{fig_semisyn_noise}{width=\textwidth,height=0.6\textheight,keepaspectratio}
{First semi-synthetic test: a synthetic gather with known first arrivals and real noise from field gather 0813. (a) Moderate and (b) strong noise, with the true first arrival (green), the conventional pick (dashed), and the FA-Agent picks (orange). (c, d) Percentage of picks within $\pm4$, $\pm8$, and $\pm20$~ms of the true first arrival for the conventional pick (gray) and FA-Agent (orange).}

\plot{fig_semisyn_weak}{width=\textwidth,height=0.47\textheight,keepaspectratio}
{Second semi-synthetic test, with weak and partly hidden far-offset first arrivals and real noise from field gather 0901. (a) The test gather displayed with a strong gain so that the noise is visible, with the true first arrival (green) and the FA-Agent picks (orange). Purple bars at the top mark the four blocks in which the first arrival was hidden when the test was built; gray bars at the bottom mark the traces removed by FA-Agent. (b) Percentage of the FA-Agent picks within $\pm4$, $\pm8$, and $\pm20$~ms of the true first arrival. (c) Decisions of FA-Agent for traces grouped by the S/N of their first arrival: for each group (number of traces $n$), the percentage of traces picked (orange) and removed (gray).}

\section{Mountainous field-data examples}
\label{sec:field}

\subsection{Data and selection of examples}

The data come from a three-dimensional land survey in a mountainous area. The data set contains 490 shots, each recorded by 28,800 channels on 40 receiver lines of 720 channels, for a total of 14,112,000 traces. Each trace has 751 samples at a 4~ms sampling interval. In this study, the shot gather is the basic unit of processing, and each example shows the 720 traces recorded on one receiver line of a shot. The agent processed all 1000 randomly sampled gathers (see Method). It first assessed the first-arrival difficulty of each gather from its display and assigned it to class A (first arrival obscured, broken, or mixed with strong interference over large parts of the gather), class B (locally weak or noisy), or class C (clear and continuous). Of these, 12 gathers were assigned to class A, 395 to class B, and 593 to class C. The gathers presented here were taken from those not used as development cases, so that they lie outside the case library of the skills. From these, gathers 0813 and 0831, which have poor data quality and class A difficulty, were selected as the main tests of performance, and the class C gather 0001 was added for comparison. Gather 0813, the gather shown in Figure~\ref{fig:fig1_0813_challenges}, is presented in detail, and gathers 0831 and 0001 are presented more briefly. The complete review of each gather by the agent is animated in the supplementary movies Mov1 (gather 0813), Mov2 (gather 0831), and Mov3 (gather 0001). For each gather the conventional initial pick was computed on the original data, and all displays use the gain described in the Method section. The agent also includes a denoising module with basic parameter tuning, which can attenuate anomalous energy before picking.

\subsection{Gather 0813}

Figure~\ref{fig:fig3_0813_review}a shows the four attention windows that the agent marked on gather 0813 after viewing the gather without and then with the initial pick. Each window corresponds to one of the difficulties in Figure~\ref{fig:fig1_0813_challenges}. Window V1 contains the strong linear event on the left and the weak first-arrival branch above it, V2 the bands of vertical noise on the left flank, V3 the apex, and V4 the far-offset zone on the right, which is covered by vertical noise and low-frequency interference. Figure~\ref{fig:fig4_0813_windows_zoom} shows an enlarged view of each window, first without picks, as the agent examined it, and then with the initial pick and the agent's decisions.

In V1 (Figure~\ref{fig:fig4_0813_windows_zoom}a and \ref{fig:fig4_0813_windows_zoom}e), the strong linear event dominates the display, but in the enlarged view a weak coherent phase can be followed above it from about trace 105. At trace 105 the initial pick lies on the strong event, about 110~ms later than the weak branch, and it reaches the weak branch only near trace 130. The agent corrected traces 105--209 onto the weak first-arrival branch. To the left of trace 105 it could not establish an earlier branch above the strong event and removed the picks.

In V2 (Figure~\ref{fig:fig4_0813_windows_zoom}b and \ref{fig:fig4_0813_windows_zoom}f), a weak ascending wavefront is visible over traces 210--248, and the agent moved the pick onto it, by up to 32~ms. Over traces 249--296, broad vertical noise and competing weak events leave no reliable wavefront on either side of a 48-trace gap, so the agent removed these picks; the wavefront reappears at trace 297, from which the initial pick was kept.

In V3 (Figure~\ref{fig:fig4_0813_windows_zoom}c and \ref{fig:fig4_0813_windows_zoom}g), the visible wavefront at the apex has local undulations, with two crests near traces 338 and 370. The initial pick is smoother than the wavefront, and the agent corrected it to follow the undulations; the changes are small, within 16~ms.

In V4 (Figure~\ref{fig:fig4_0813_windows_zoom}d and \ref{fig:fig4_0813_windows_zoom}h), the wavefront remains visible to about trace 562, and the agent corrected this short continuation. Beyond it, vertical noise and low-frequency interference cover the zone where the first arrival is expected. Coherent events are visible below the interference, but they are later arrivals and do not establish the first arrival. The agent therefore removed the far-offset picks, whereas the initial pick continues through the interference.

The decisions for the whole gather are shown in Figure~\ref{fig:fig3_0813_review}b. Between the windows, the initial pick lies on a clear wavefront and was kept. Seven short interruptions of a wavefront visible on both sides, 39 traces in total, were bridged. Overall, 103 picks were kept, 307 corrected, and 310 removed, leaving picks on 410 of the 720 traces (Table~\ref{tbl:field}). Mov1 shows this review as a continuous sequence. The initial pick is drawn across the gather, and a scan over the whole gather marks windows V1--V4 as it passes them. Each window is then enlarged, viewed without and with the picks, and corrected, and the final picks are displayed on the whole gather. The movie shows how each decision follows from what the agent sees.

\subsection{Gathers 0831 and 0001}

Gather 0831 (Figure~\ref{fig:fig5_0831_review}a) shows a similar pattern of difficulties. On the left (V1), the initial pick follows a strong linear event with no earlier wavefront visible above it, and the agent removed these picks. At the apex (V3), the continuity constraint had flattened the initial pick into a plateau at 224~ms, and the agent replaced it with the curved visible wavefront, which reaches about 176~ms. At far offsets (V4), a weak wavefront was followed to trace 665 and the remaining traces were removed. Four dead traces received no pick. The most informative window is V2 (Figure~\ref{fig:fig5_0831_review}b and \ref{fig:fig5_0831_review}c), where a broad band of vertical noise crosses the left flank. In the enlarged view, a weak oblique wavefront can be followed above the strong later wave trains from about trace 195 to 237, whereas the initial pick runs below it, through the noise and onto the later events. The agent corrected traces 195--237 onto this wavefront, moving the picks earlier by 44~ms on average and up to 96~ms, and removed traces 141--194, where the wavefront could not be followed through the noise band. The full review of this gather is shown in Mov2.

Gather 0001 (Figure~\ref{fig:fig6_0001_review}a) has a clearer first arrival over most of its length. The agent kept the initial pick on the left flank, where it lies on the visible wavefront, and made local corrections where a band of vertical noise crosses the wavefront (V1) and along the irregular crest (V2). At far offsets (V4), a strong later wave train is visible, but no earlier branch could be confirmed, so the picks were removed; this zone includes five dead traces. The most informative window is the right flank V3 (Figure~\ref{fig:fig6_0001_review}b and \ref{fig:fig6_0001_review}c). In the enlarged view, a weak wavefront runs above the stronger phases on which the initial pick lies. The agent moved the picks of traces 440--580 onto this earlier wavefront, by 47~ms on average and up to 132~ms, and removed traces 581--615, where the earlier wavefront could no longer be followed. The full review of this gather is shown in Mov3.

\subsection{Summary of the examples}

Table~\ref{tbl:field} summarizes the decisions. The proportion of removed traces follows the difficulty of the gathers: 43\% of the traces were removed in 0813, 35\% in 0831, and 19\% in 0001. In all three gathers the removals are concentrated in the zones identified in the whole-gather recognition, mainly at far offsets and under strong interference, whereas the initial pick was kept wherever it lay on a visible wavefront. Corrections were made in every window that contained a visible wavefront differing from the initial pick.

\plot{fig3_0813_review}{width=\textwidth,height=0.68\textheight,keepaspectratio}
{Agent review of gather 0813. (a) Initial pick (green) and the attention windows V1--V4 marked by the agent. (b) Final picks: kept (green), corrected (orange), and removed (blank); the dashed line is the initial pick, and the bar above the panel shows the decision for each trace.}

\plot{fig4_0813_windows_zoom}{width=\textwidth,height=0.68\textheight,keepaspectratio}
{Enlarged views of the attention windows of gather 0813. Top row (a--d): windows V1--V4 without picks, as examined by the agent. Bottom row (e--h): the same views with the initial pick (dashed) and the agent's decisions: kept (green), corrected (orange), and removed (gray bar at the bottom).}

\plot{fig5_0831_review}{width=\textwidth,height=0.68\textheight,keepaspectratio}
{Agent review of gather 0831. (a) Initial pick (dashed), attention windows, and final picks, displayed as in Figure~\ref{fig:fig3_0813_review}b; the dashed box marks the enlarged view. (b, c) Enlarged view of window V2 without picks and with the initial pick and the agent's decisions.}

\plot{fig6_0001_review}{width=\textwidth,height=0.68\textheight,keepaspectratio}
{Agent review of gather 0001, displayed as in Figure~\ref{fig:fig5_0831_review}. (b, c) Enlarged view of the right-flank window V3 without picks and with the initial pick and the agent's decisions.}

\section{Discussion and future work}
\label{sec:discussion}

\subsection{Visual recognition and the reliability of picks}

Visual recognition is the central element of FA-Agent. Conventional pickers based on energy ratios and related attributes respond to the strongest or most abrupt change in a trace. In mountainous data they therefore often pick a later, more energetic arrival or a noise burst. Deep-learning pickers learn the arrival patterns of the surveys on which they were trained. On a new survey with different acquisition, waveforms, and noise, they usually require retraining or fine-tuning with labeled data from that survey \citep{Mardan2024,LiH2024}; without it, they generalize poorly \citep{Wang2024}. First-arrival tomography, however, requires accuracy rather than completeness. An erroneous pick does more harm than a missing one, so a trace is better left blank than given a wrong pick. The visual approach meets this requirement by emulating how a processing geophysicist works: it looks at the gather, judges which event is the first arrival, and keeps a pick only where the arrival can be seen. The expert experience behind these judgments is written into the skills, so the agent applies it to every gather in the same way and needs no retraining. Expert picks in such data are themselves subjective, so we verified this behavior on semi-synthetic gathers with known first arrivals, where the agent kept accurate picks and removed those whose first arrival had been hidden (Figures~\ref{fig:fig_semisyn_noise} and \ref{fig:fig_semisyn_weak}).

\subsection{A framework complementary to existing pickers}

FA-Agent complements conventional and deep-learning pickers rather than replacing them. The present implementation calls a single conventional picker for the initial pick. Any other picker, including a trained neural network, can be added as a tool, and the agent then decides from the gather display where its picks can be trusted. What we provide is a framework: a multimodal model that recognizes and decides, tools that it can call, and skills that record how the task is done. Others can build on it by embedding their own picking, denoising, or quality-control tools and by adding skills for their data.

\subsection{Computational cost and token consumption}

The correction of one gather took about 30--90~s on a single core of a local AMD CPU, depending on the difficulty of the gather. This time should decrease as the skills are streamlined and as large language models become faster. A commercial large language model also consumes tokens, mainly for image input, which adds cost when many gathers are processed. In addition, seismic data are often confidential, and sending them to an external service may not be acceptable. A locally deployed multimodal large language model addresses both concerns and is a natural choice for future applications.

\subsection{Accumulating experience}

Because experience is stored in the skills rather than in the model weights, FA-Agent can keep improving with use. When an expert reviews a processed gather, the agent writes the gather into the case library of the skills, and later decisions can draw on more cases. In this sense the agent becomes more experienced, and smarter, the more it is used. The same holds when others use FA-Agent: each user who applies it to new data and reviews the results adds cases and refines the skills, so the agent becomes progressively more complete.

\subsection{Model choice and local deployment}

FA-Agent is not tied to GPT-6 Astra. Any multimodal large language model that can view images and call tools can serve as its decision maker. The visual capability of such models continues to improve, which should help the agent recognize weak arrivals. Locally deployed models are particularly suitable because they keep the data on site and avoid per-call costs. Future work will apply FA-Agent to further surveys, extend its set of tools, and use locally deployed models for routine processing.

\section{Conclusions}
\label{sec:conclusions}

We have developed FA-Agent, an agent framework for first-arrival picking on difficult data. A multimodal large language model performs visual recognition and makes decisions, tools carry out the deterministic processing, and skills hold the processing experience; together they form an open framework that works like a software package. FA-Agent follows the workflow of a seismic data processor: it surveys the whole gather to identify the zones that need attention, examines them in enlarged views, and keeps, corrects, or removes the picks. In this way, visual recognition informed by expert experience replaces picking that relies only on energy attributes or on training data. On semi-synthetic gathers with known first arrivals, the agent picked visible first arrivals accurately and removed hidden ones instead of picking later events. On mountainous data with a low signal-to-noise ratio, the agent recognized weak first arrivals masked by strong direct waves and noise and corrected the picks onto them. It removed the picks in zones where no first arrival could be confirmed, so the picks it kept are reliable for first-arrival tomography, also in the most difficult gathers. FA-Agent complements existing pickers, and any picking, denoising, or quality-control tool can be embedded in it. Its experience grows through the skills as it is used, and it can run on locally deployed models. FA-Agent therefore offers a practical path toward automating first-arrival picking in routine processing, in which experts review the agent's results instead of picking every gather by hand.

\begin{acknowledgments}
We thank a geophysicist with many years of experience in processing mountainous seismic data whose discussions on first-arrival picking shaped the skills of FA-Agent. We also thank the Earth Observatory of Singapore and the School of Civil and Environmental Engineering, Nanyang Technological University, for supporting this research.
\end{acknowledgments}

\section*{Data and materials availability}

The field data used in this study are confidential and cannot be released. The complete code of the FA-Agent framework, including its tools and skills, will be made publicly available after the initial review of this paper, together with example data for testing.

\section*{Supplementary material}

The movies are provided as ancillary files of this arXiv submission.

\begin{description}
\item[Mov1.] Agent review of gather 0813. The movie shows, in sequence, the input gather, the conventional initial pick, the whole-gather scan in which the attention windows V1--V4 are marked, the enlarged inspection and correction of each window, and the final picks on the whole gather. Green: initial or kept picks; orange: corrected picks; removed picks are left blank.
\item[Mov2.] Agent review of gather 0831, displayed as in Mov1.
\item[Mov3.] Agent review of gather 0001, displayed as in Mov1.
\end{description}

\bibliographystyle{seg}
\bibliography{references}
\clearpage
\begin{table}[H]
\centering
\caption{Decisions of FA-Agent on the three field gathers (720 traces each). Corrected traces include anchors that coincide with the initial pick; the number of traces whose traveltime changed is given in parentheses. Bridged traces are short gaps interpolated between visible wavefronts on both sides.}
\label{tbl:field}
\begin{tabular}{lcccccc}
\hline
Gather & Windows & Kept & Corrected & Removed & Traces with picks & Bridged \\
\hline
0813 & 4 & 103 & 307 (268) & 310 & 410 (57\%) & 39 \\
0831 & 4 & 255 & 215 (199) & 250 & 470 (65\%) & 25 \\
0001 & 4 & 278 & 302 (259) & 140 & 580 (81\%) & 14 \\
\hline
\end{tabular}
\end{table}

\end{document}